\documentclass[prd,showpacs,preprintnumbers,nofootinbib,aps,9pt]{revtex4}
\usepackage{subeqnarray}
\usepackage{epsfig,amsmath,amssymb}
\usepackage{mathrsfs}
\usepackage[usenames,dvipsnames]{color}
\usepackage[pagebackref=true, colorlinks=true]{hyperref}
\definecolor{redish}{rgb}{0.7,0.2,0.0}  
\definecolor{bluish}{rgb}{0.2,0.5,0.8}
\hypersetup{linkcolor=redish,          
                  citecolor=blue,        
                  filecolor=magenta,      
                  urlcolor=bluish}          
\newcommand{\ba}{\nopagebreak[3]\begin{eqnarray}}
\newcommand{\ea}{\end{eqnarray}}
\newcommand{\bii}{\begin{itemize}}
\newcommand{\eii}{\end{itemize}}
\newcommand{\nn}{\nonumber}
\newcommand{\f}{\frac}
\def \pa{\partial}
\def \d{~\delta}
\def \a{\alpha}
\def \b{\beta}
\def \s{\sigma}
\def \l{\ell}
\def \i{\zeta}
\def \e{\epsilon}

\def \pr{\prime}
\def \({\left(}
\def \){\right)}
\def \[{\left[}
\def \]{\right]}
\def \g{\gamma}
\def \f{\frac}
\def \th{\theta}
\def \Th{\Theta}
\def \Ra{\Rightarrow}
\def \p{\phi}
\def \P{\Phi}
\def \G{\Gamma}
\def \D {\Delta}
\def \Si {\Xi}
\def \lm {\lambda}

\def \heq{\hat{=}}
\def \cd{\nabla}
\def \Li{{\cal L}}
\def\pb#1{\rlap{\lower1.5ex\hbox{$\longleftarrow$}}{#1}}
\def\dpb#1{\rlap{\lower1.5ex\hbox{$\Longleftarrow$}}{#1}}
\def\spb#1{\rlap{\lower1.5ex\hbox{$\leftarrow$}}{#1}}
\def\sdpb#1{\rlap{\lower1.5ex\hbox{$\Leftarrow$}}{#1}}
\def \A{A^{\prime}}
\def \B{B^{\prime}}
\def \C{C^{\prime}}
\begin{document}
\title{Field dynamics on the trapping horizon in Vaidya spacetime}
\author{Abhishek Majhi}
\email{abhim@imsc.res.in}

\affiliation{Instituto de Ciencias Nucleares\\
Universidad Nacional Autonoma de Mexico\\
A. Postal 70-543, Mexico D.F. 04510, Mexico\\}

\affiliation{The Institute of Mathematical Sciences\\4th Cross St., CIT Campus, Taramani,\\Chennai, Tamil Nadu, India\\}

\affiliation{Astro-Particle Physics and Cosmology Division\\Saha Institute of Nuclear Physics\\Kolkata-700064, India}%

\pacs{}
\begin{abstract}
In this article, we shed some light on the field theoretic aspect of the spherically symmetric trapping horizon in Vaidya spacetime. The effective field equations are that of a Chern-Simons theory coupled to bulk sources through two different couplings, one is purely geometric and the other is matter dependent. This is an effective generalization of the equilibrium horizon scenario where the Chern-Simons theory is coupled to the bulk geometry through a constant matter-independent coupling. Further, we note that the field equations pulled-back to a cross-section of the horizon is inadequate to manifest the nature of the horizon. Hence, contrary to the usual practice, the evolution equation needs to be considered at least while passing on to the quantum theory.

\end{abstract}
\maketitle

\section{Introduction}
A black hole horizon in gravitational as well as thermodynamic equilibrium is effectively described by an {\bf isolated horizon(IH)}\cite{ih1,ih2}. At the formulation level, an IH is considered to be a 2+1 dimensional null inner boundary of 3+1 dimensional spacetime satisfying certain boundary conditions {\it on the } IH so as to capture all the properties of an event horizon without having to impose any restriction, such as stationarity, on the bulk spacetime off the horizon\cite{ih1}.  In Hayward's language, an IH is a null future outer trapping horizon\cite{hay1}. Formulation of gravity as a gauge theory reveals that the differentiability of the action of a spacetime with IH as inner boundary needs the presence of a 2+1 dimensional topological {\bf Chern-Simons(CS)} action on the IH with a coupling proportional to its area\cite{ih2}. In the context of the present work, it is to be noted that  the boundary conditions defining the IH ensures that there are no matter degrees of freedom on the IH.

 On the other hand, a  non equilibrium black hole horizon is effectively described by a non-null future outer trapping  horizon\cite{hay1}\footnote{According to Ashtekar and Krishnan, a spacelike future outer trapping horizon is called a dynamical horizon which represnts a growing black hole\cite{dh1}. However, the timelike future outer trapping horizon is not considered as a black hole horizon since it represents a shrinking horizon. Hence, we do not use the word `dynamical horizon' here as we shall approach the problem so as to tackle all the three situations i.e. null, spacelike and timelike.}.  In this work, we shall use the concept of {\bf trapping horizon(TH)} so as to take into account all the possibilities i.e. null, spacelike and timelike horizons. A common example of a spacetime admitting a TH is the Vaidya spacetime\cite{vaidya} which becomes null TH or IH when the horizon mass becomes constant i.e. the spacetime becomes that of a Schwarzschild black hole admitting the event horizon which is a special case of IH. Having known the field theory on an equilibrium black hole horizon, it will be interesting to see how the field equations on the null TH, that leads to the CS theory, gets modified in the case of a non-equilibrium TH which allows matter and radiation to cross it.

  As a first step of this program, we shall consider the case of Vaidya spacetime which admits a {\bf spherically symmetric} TH({\bf SSTH}).  In terms of the {\bf Newman-Penrose(NP)} coefficients, in the spherically symmetric case the component $\P_{00}$(to be defined later), evaluated on the TH, carries the all needed information which determines whether the SSTH is null or not \cite{hay2}.  In other words it can be said that $\P_{00}$ on an SSTH alone suffices to determine whether it is in equilibrium or out of equilibrium. This makes the exercise mathematically easier to handle, but still enables us to shed some light on the field theoretic aspect  of an SSTH. The general scenario, where the other NP coefficients will also dictate the nature of the TH\cite{booth,rezzolla,jaramillo}, is kept for future study.


Let us debrief the structure of the paper. In section II, we shall discuss the methodology that leads to the derivation of the CS boundary term in case of IH. In section III, we shall lay down the overall framework by discussing the field equations on a three dimensional hypersurface in the NP formalism. Then, in section IV, we shall apply these equations to the SSTH in the Vaidya and Schwarzschild spacetime. Finally, we end with a few concluding remarks in section V.



\section{Chern-Simons term on IH in a nutshell}
Let us consider a three dimensional hypersurface in four dimensional spacetime, which may be timelike, spacelike or null and is foliated by 2-spheres$(S^2)$ i.e. topologically $S^2\times R$.  Further, we consider that the hypersurface and hence, each $S^2$ is adapted to a set of null tetrads $(\l, n, m, \bar m)$ with $\l$ and $n$ normal to the spacelike 2-sphere $S^2$. For convenience and to compare the scenario with the original formulation of the CS theory in IH case in \cite{ih2}, we shall follow the same spinorial language throughout our analysis. Hence, the basic structures will be alike, but the analysis will be generalized to the SSTH in Vaidya spacetime rather than to only IH.

The basic fields are the soldering forms $\s_a^{~AA^{\pr}}$ for the SL(2,C) (primed and unprimed) spinors and  the SL(2,C) connections $A_{a}^{~AB}$ on unprimed spinors\footnote{For details on these issues one can look into \cite{ih2} and the references there in.}. The variation of the gravitational action with respect to the connection variable $A$, so as to have field equations satisfied in the bulk, yields the boundary term given by\cite{ih1,ih2}
\ba
 \d S_{grav}|_{bdy}=-\f{i}{8\pi}\int_{bdy}\text{Tr}~ \Si\wedge \d A\label{action}
\ea
where we have set the gravitational constant to unity, `Tr' denotes the trace over the internal indices, `bdy' stands for boundary and  $\Si_{ab}^{~~AB}$ is a self-dual two form field constructed of the soldering forms given by
\ba
\Si_{ab}^{~~AB}=2\e_{A^{\pr}C^{\pr}}\s_{[a}^{AA^{\pr}}\s_{b]}^{BC^{\pr}}
=2\s_{[a}^{AA^{\pr}}\s_{b]A^{\pr}}^{B}\label{sol}
\ea 
Now, the self-dual part, denoted by {\bf superscript `$+$' }, of the Riemann tensor is given by
\ba
R_{ab}^{+~AB}=-\f{1}{4}\Si_{ab}^{~~AB}R_{cd}^{~~ab}\label{key}
\ea
This is the starting equation for the derivation of the CS theory in case of IH\cite{ih2}.
Taking the pullback of the above equation and expressing the Riemann tensor in terms of the NP coefficients, the boundary conditions of the IH yields the simple relation $\text{Tr}~\spb{\Si}\wedge\pb{\d A}~\heq-\f{a_{IH}}{2\pi}~\text{Tr}~\spb{R^+}\wedge\pb{\d A}$, $a_{IH}$ being the area of the cross sections of the IH which remains constant along the IH. Using this relation directly in eq.(\ref{action}) one obtains on the right hand side precisely the variation of the CS action.

\section{Field equations on a three dimensional hypersurface}
While generalizing the analysis to an SSTH one would expect that the relation may not be so simple any more because most of the boundary conditions will not be valid, except the ones to ensure the light-trapping property and the spherical symmetry of the TH.
Here we shall calculate the pullback of eq.(\ref{key}) on an arbitrary three surface which is topologically $S^2\times R$ and then apply those equations in case of the SSTH in Vaidya spacetime.  Hence, this work can be regarded as a simple minded first step towards finding the field theory on a generic TH. The general analysis for a generic TH will be reported in some future publication.
\par
Let us begin our analysis as follows. The null tetrads adapted to an $S^2\times R$ three dimensional hypersurface are defined in terms of spinors and the soldering forms as follows
\ba
\l^a&=&~i o^Ao^{A^{\prime}}\s^a_{A{A^{\prime}}}\nn\\
n^a&=&~i \i^A\i^{A^{\prime}}\s^a_{A{A^{\prime}}}\nn\\
m^a&=&~o^A\i^{A^{\prime}}\s^a_{A{A^{\prime}}}\nn\\
\bar m^a&=&-\i^Ao^{A^{\prime}}\s^a_{A{A^{\prime}}}\nn
\ea
where $(o^A,\i^A)$ form a spinor dyad  satisfying the following relations: 
\ba
\e_{AB}\i^Ao^B=\i^Ao_A=-\e_{BA}\i^Ao^B=-\i_Bo^B=1,
\ea 
$\e_{AB}$ being the antisymmetric 2-form on the spinor space. Further, it will be considered throughout the text that 
\ba
\e_{AB}\i^B=-\e_{BA}\i^B=\i_A~~\text{and}~~\e_{AB}o^B
=-\e_{BA}o^B=o_A.
\ea
The induced metric on $S^2$ is given by 
\ba
\tilde{q}_{ab}=g_{ab}+2\l_{(a}n_{b)}
\ea
where $g_{ab}$ is the full spacetime metric which can be written in terms of the soldering forms as 
\ba
g_{ab}=\e_{AB}\e_{A^{\pr}B^{\pr}}\s_a^{AA^{\pr}}\s_b^{BB^{\pr}}.
\ea 
Now, the solder forms can be written in terms of the spinor dyads and the null tetrads as
\ba
\s^a_{AA^{\pr}}=-i(\i_A\i_{A^{\pr}}\l^a+o_Ao_{A^{\pr}}n^a)-\i_Ao_{A^{\pr}}m^a+o_A\i_{A^{\pr}}\bar m^a
\ea
A bit of spinorial algebra leads to the following results
\ba
\i^Ao_{\C}\s_b^{B\C}&=&-i\i^A\i^B\l_b+\i^Ao^B\bar m_b\nn\\
o^A\i_{\C}\s_b^{B\C}&=&io^Ao^B n_b-o^A\i^Bm_b\nn\\
\i^A\i_{\C}\s_b^{B\C}&=&i\i^Ao^B n_b-\i^A\i^Bm_b\nn\\
o^Ao_{\C}\s_b^{B\C}&=&-io^A\i^B\l_b+o^Ao^B\bar m_b\nn
\ea
Using the above results in eq.(\ref{sol}) the self-dual two-form field $\Si$ can be calculated to be
\ba
\Si_{ab}^{~~AB}=4i\i^A\i^B\l_{[a}m_{b]}+4io^Ao^Bn_{[a}\bar m_{b]} +4\i^{(A}o^{B)}m_{[a}\bar m_{b]}+4\i^{(A}o^{B)}n_{[a}\l_{b]}\label{si}
\ea 
Now, the Riemann tensor components in NP formalism can be written in terms of the Ricci and Weyl scalars and are given by the following two spinorial equations
\ba
\P_{AB\A\B}&=&\P_{00}\i_A\i_B\i_{\A}\i_{\B}-2\P_{01}\i_A\i_B\i_{(\A}o_{\B)}
+\P_{02}\i_A\i_Bo_{\A}o_{\B}\nn\\
&&~~~-2\P_{10}\i_{(A}o_{B)}\i_{\A}\i_{\B}
+4\P_{11}\i_{(A}o_{B)}\i_{(\A}o_{\B)}-2\P_{12}\i_{(A}o_{B)}o_{\A}o_{\B}\nn\\
&&~~~~~~~~~~~~~~+\P_{20}o_Ao_B\i_{\A}\i_{\B}-2\P_{21}o_Ao_B\i_{(\A}o_{\B)}
+\P_{22}o_Ao_Bo_{\A}o_{\B}\label{riem1}
\ea
and
\ba
\Psi_{ABCD}&=&\Psi_0\i_A\i_B\i_C\i_D-4\Psi_1\i_{(A}\i_{B}\i_{C}o_{D)}+6\Psi_2\i_{(A}\i_{B}o_{C}o_{D)}
-4\Psi_3\i_{(A}o_{B}o_{C}o_{D)}+\Psi_4o_Ao_Bo_Co_D\label{riem2}
\ea
where we shall consider the definitions of the NP coefficients in \cite{ih2}.
Using eq.(\ref{riem1}) and eq.(\ref{riem2}) and writing the Riemann tensor in terms of the NP coefficients and the Ricci curvature scalar $R$, eq.(\ref{key}) can be explicitly written as follows
\ba
R^+_{abCD}&=&-\f{1}{2}\Psi_{ABCD}\Si_{ab}^{~~AB}-\f{1}{2}\bar\P_{\A\B CD}\bar{\Si}_{ab}^{~~\A\B}-\f{R}{24}\Si_{abCD}\nn\\
&~&~\nn\\
&=&-2i\l_{[a}m_{b]}\[\i_C\i_D\(\Psi_2+\f{R}{12}\)-2\i_{(C}o_{D)}\Psi_3+o_Co_D\Psi_4\]\nn\\
&~&~\nn\\
&~&~~-2in_{[a}\bar 
m_{b]}\[\i_C\i_D\Psi_0-2\i_{(C}o_{D)}\Psi_1+o_Co_D\(\Psi_2+\f{R}{12}\)\]\nn\\
&~&~\nn\\
&~&~~~~-2m_{[a}\bar m_{b]}\[\i_C\i_D(\Psi_1-\bar{\P}_{10})-2\i_{(C}o_{D)}\(\Psi_2-\bar{\P}_{11}-\f{R}{24}\)-o_Co_D\bar{\P}_{12}\]\nn\\
&~&~\nn\\
&~&~~~~~-2n_{[a}\l_{b]}\[\i_C\i_D(\Psi_1+\bar{\P}_{10})-2\i_{(C}o_{D)}\(\Psi_2+\bar{\P}_{11}-\f{R}{24}\)+o_Co_D\bar{\P}_{12}\]\nn\\
&~&~\nn\\
&~&~~~~~~~~+2in_{[a}m_{b]}\[\i_C\i_D\bar{\P}_{00}-
2\i_{(C}o_{D)}\bar{\P}_{01}+o_Co_D\bar{\P}_{02}\]\nn\\
&~&~\nn\\
&~&~~~~~~~~~~~~+2i\l_{[a}\bar m_{b]}\[\i_C\i_D\bar{\P}_{20}-2\i_{(C}o_{D)}\bar{\P}_{21}+o_Co_D
\bar{\P}_{22}\]\label{full}
\ea
\vspace{1cm}

{\bf Calculating the pullbacks of $R^+$ and $\Si$:} Now, for every choice of $\l$ and $n$ we can define a scalar field $\mu$ on the hypersurface\cite{hay1} such that $h^a=\l^a-\mu n^a$ and $k^a=\l^a+\mu n^a$, where $h$ is the vector field along the hypersurface and $k$ is the vector field transverse to $h$ i.e. normal to the hypersurface. Hence, $h.h\heq2\mu$ and $k.k\heq-2\mu$. It is quite easy to see from the norm of $h$ that the scalar field $\mu$ alone suffices to manifest whether the hypersurface is null ($\mu=0$), spacelike ($\mu>0$) or timelike ($\mu<0$)\footnote{We consider the signature of the spacetime metric to be $(-,+,+,+)$ all the way through.}. 

While taking the pullback of eq.(\ref{si}) and eq.(\ref{full}) on the hypersurface, we have $\spb{k_a}\heq 0$ because the vector field $k$ is normal to the hypersurface. Hence we have the following two cases:
\begin{itemize}
\item For {\it null or degenerate} hypersurface i.e. $\mu=0$, it leads to $\spb{\l_a}\heq0$. 
\item For {\it non-null or non-degenerate} hypersurface i.e. $\mu\neq 0$, it leads to $\spb{\l_a}\heq -\mu\spb{n_a}$ and $\spb{h_a}\heq -2\mu\spb{n_a}$.
\end{itemize}
Thus it is quite trivial to see that $n\pb{_{[a}\l_{b]}}\heq 0$ irrespective of the nature of the hypersurface. Hence, the pullback of $\Si$ on the hypersurface is given by
\ba
\pb{\Si_{ab}}^{AB}=4i\i^A\i^B\l\pb{_{[a}m_{b]}}+4io^Ao^Bn\pb{_{[a}\bar m_{b]}} +4\i^{(A}o^{B)}m\pb{_{[a}\bar m_{b]}}\label{sipb}
\ea
which for null hypersurface i.e. $\mu=0$ can be recast as
\ba
\pb{\Si_{ab}}^{AB}=4io^Ao^Bn\pb{_{[a}\bar m_{b]}} +4\i^{(A}o^{B)}m\pb{_{[a}\bar m_{b]}}
\ea
and for non-null hypersurface i.e. $\mu\neq 0$ can be rewritten as
\ba
\pb{\Si_{ab}}^{AB}=2i\i^A\i^Bh\pb{_{[a}m_{b]}}-\f{2}{\mu}io^Ao^Bh\pb{_{[a}\bar m_{b]}} +4\i^{(A}o^{B)}m\pb{_{[a}\bar m_{b]}}\label{sigth}
\ea
Considering that $n\pb{_{[a}\l_{b]}}\heq~0$  irrespective of the nature of the hypersurface and noting that $\bar{\P}_{00}=\P_{00}, ~\bar{\P}_{11}=\P_{11},~ \bar{\P}_{22}=\P_{22}$, we can write pullback of eq.(\ref{full}) for an SSTH as follows
\ba
\pb{R^+_{abCD}}
&\heq&-2i\l\pb{_{[a}m_{b]}}\[\i_C\i_D\(\Psi_2+\f{R}{12}\)-2\i_{(C}o_{D)}\Psi_3+o_Co_D\Psi_4\]\nn\\
&~&~\nn\\
&~&~~-2in\pb{_{[a}\bar 
m_{b]}}\[\i_C\i_D\Psi_0-2\i_{(C}o_{D)}\Psi_1+o_Co_D\(\Psi_2+\f{R}{12}\)\]\nn\\
&~&~\nn\\
&~&~~~~-2m\pb{_{[a}\bar m_{b]}}\[\i_C\i_D(\Psi_1-\bar{\P}_{10})-2\i_{(C}o_{D)}\(\Psi_2-{\P}_{11}-\f{R}{24}\)-o_Co_D\bar{\P}_{12}\]\nn\\
&~&~\nn\\
&~&~~~~~~~~+2in\pb{_{[a}m_{b]}}\[\i_C\i_D{\P}_{00}-
2\i_{(C}o_{D)}\bar{\P}_{01}+o_Co_D\bar{\P}_{02}\]\nn\\
&~&~\nn\\
&~&~~~~~~~~~~~~+2i\l\pb{_{[a}\bar m_{b]}}\[\i_C\i_D\bar{\P}_{20}-2\i_{(C}o_{D)}\bar{\P}_{21}+o_Co_D
{\P}_{22}\]\label{fpb}
\ea
which for a null hypersurface i.e. $\mu=0$ can be recast as 
\ba
\pb{R^+_{abCD}}
&\heq&-2in\pb{_{[a}\bar 
m_{b]}}\[\i_C\i_D\Psi_0-2\i_{(C}o_{D)}\Psi_1+o_Co_D\(\Psi_2+\f{R}{12}\)\]\nn\\
&~&~\nn\\
&~&~~~~-2m\pb{_{[a}\bar m_{b]}}\[\i_C\i_D(\Psi_1-\bar{\P}_{10})-2\i_{(C}o_{D)}\(\Psi_2-{\P}_{11}-\f{R}{24}\)-o_Co_D\bar{\P}_{12}\]\nn\\
&~&~\nn\\
&~&~~~~~~~~+2in\pb{_{[a}m_{b]}}\[\i_C\i_D{\P}_{00}-
2\i_{(C}o_{D)}\bar{\P}_{01}+o_Co_D\bar{\P}_{02}\]\label{fpbn}
\ea
and for non-null hypersurface i.e. $\mu\neq 0$ can be rewritten as
\ba
\pb{R^+_{abCD}}
&\heq&-ih\pb{_{[a}m_{b]}}\[\i_C\i_D\(\Psi_2+\f{R}{12}+\f{{\P}_{00}}{\mu~~}\)-2\i_{(C}o_{D)}\(\Psi_3+\f{\bar{\P}_{01}}{\mu~~}\)+o_Co_D\(\Psi_4+\f{\bar{\P}_{02}}{\mu~~}\)\]\nn\\
&~&~\nn\\
&~&~~-2m\pb{_{[a}\bar m_{b]}}\[\i_C\i_D(\Psi_1-\bar{\P}_{10})-2\i_{(C}o_{D)}\(\Psi_2-{\P}_{11}-\f{R}{24}\)-o_Co_D\bar{\P}_{12}\]\nn\\
&~&~\nn\\
&~&~~~~~+ih\pb{_{[a}\bar m_{b]}}\[\i_C\i_D\(\bar{\P}_{20}+\f{\Psi_0}{\mu~}\)-2\i_{(C}o_{D)}\(\bar{\P}_{21}+\f{\Psi_1}{\mu~}\)+o_Co_D
\{{\P}_{22}+\f{1}{\mu}\(\Psi_2+\f{R}{12}\)\}\]~~~~~~~\label{fpbnn}
\ea
The equations showing the pullbacks of $R^+$ and $\Si$ are quite messy. So, we shall focus on a scenario which makes those equations simpler. As it was pointed earlier that $\Phi_{00}$ is the only NP coefficient which dictates the nature of an SSTH, it would have been nice to study the these formidable equations in some simple spacetime admitting an SSTH such that only $\Phi_{00}$ is the non vanishing component among the Ricci scalars. Fortunately, we have such a spacetime at our disposal, namely, the Vaidya spacetime. We shall study the above equations of the pullbacks of the relevant fields in case of the SSTH in Vaidya spacetime and also see how it can be compared with the situation when the SSTH becomes null, whose field equations are already known to us.

\section{Field equations on SSTH in Vaidya and Schwarzschild spacetime}
 Now, let us study the pullbacks $\spb{R^+}$ and $\spb{\Si}$ on the SSTH admitted by the Vaidya spacetime given by the following metric
 \ba
ds^2=-\(1-\f{2M(v)}{r}\) dv^2 + 2 dv dr + r^2 d\th ^2 + r^2\sin ^2\th ~d\phi ^2\label{vm}\nn
\ea
where $v$ is the ingoing null coordinate and $M(v)$ is the coordinate dependent mass associated with the horizon. The metric is a spherically symmetric solution of the Einstein's equations with energy momentum tensor of a null-dust given by $T_{ab}=\f{\dot{M}}{4\pi r^2}\cd_av\cd_bv$, where $\dot M\equiv dM(v)/dv$. For $\dot M=0$ i.e. constant horizon mass, the energy-momentum tensor vanishes and the metric reduces to the spherically symmetric vacuum solution of the Einstein's equations i.e. the Schwarzschild metric. Here, one can check that the hypersurface $\Phi\equiv r-2M(v)=0$ is  {\it non-degenerate}, i.e. $g^{ab}(\pa_a\Phi)(\pa_b\Phi)\heq -4\dot M\neq 0$ for $\dot M \neq 0$, where $g^{ab}$ is the inverse of the metric whose nonzero components are $g^{rr}=\D, ~~g^{rv}=g^{vr}=1,~~g^{\th \th}=r^{-2},~~g^{\p \p}=r^{-2}\sin^{-2}\th$.   It is not hard to see that for {\it constant} horizon mass ($\dot M=0$), the hypersurface $\Phi\equiv r-2M=0$ becomes {\it null} or degenerate, the energy momentum tensor vanishes and the metric is the spherically symmetric vacuum solution of the Einstein's equations, namely, the Schwarzschild solution. 
For Vaidya spacetime, the non-vanishing NP coefficients on the SSTH are $\P_{00}, \Psi_2$ and $\Psi_4$, of which $\Psi_4$ will not be required in our calculations, and the others vanish. The most important one is obviously $\Phi_{00}$, only which, among the three, depends on $\dot M$. One can calculate to find that
\ba
\Phi_{00}&:=&\f{1}{2}R_{ab}\ell^a\ell^b~\heq~ \dot M/4M^2\label{p00}\\
\Psi_2&:=&C_{abcd}\ell^am^b\bar m^cn^d~\heq~-1/8M^2\label{si2}
\ea
where $R_{ab}$ is the Ricci tensor and $C_{abcd}$ is the Weyl tensor, according to the definitions used in \cite{ih2}. Thus it is clearly visible that for $\dot M=0$, $\Phi_{00}$, as well as $\mu$(according to our proposal), vanish, which indicate that the hypersurface $\Phi=0$ becomes null and  thus  becomes the event horizon of the Schwarzschild black hole spacetime.

To show that the hypersurface $\Phi=0$ is a TH, one has to calculate the expansions of the outward directed null vector field on the 2-spheres which are the leaves of foliation of the TH. The expansion of a 2-sphere along a vector field $X$ is defined as $\Th_{(X)}=q^{ab}\cd_aX_b$, where $q^{ab}$  is the inverse of the induced 2-metric $q_{ab}$ on the 2-sphere which is given by $q_{ab}=r^2\cd_a\th\cd_b\th+r^2\sin^2\th\cd_a\p\cd_b\p$. One can check the following conditions are satisfied on the horizon defined by $\Phi=0$ both for Vaidya $(\dot M\neq 0)$ and Schwarzschild $(\dot M=0)$ case $\Th_{(\ell)}=\f{r-2M(v)}{r^2}~\heq~ 0$ i.e. the expansion of the future directed null normal vanishes for each $S^2$ on the hypersurface$(\Phi=0)$. To check this, one can show that the {\it nonzero} Christoffel symbols corresponding to the metric (\ref{vm}) are :
\ba
&&\G^{v}_{vv}=\f{M}{r^2}~,~~~~~\G^{v}_{\th \th}=-r ~,~~~
\G^{v}_{\p \p}=-r\sin^2\th ~,~~~~~~
\G^{r}_{vv}=\f{\D M}{r^2}+\f{\dot M}{r}~~, \nn\\
&&\G^{r}_{vr}=\G^{r}_{rv}=-\f{M}{r^2}~,~~~~~~~
\G^r_{\th\th}=-\D r~,~~~\G^r_{\p\p}=-\D r \sin^2\th~,\nn\\
&&\G^{\th}_{r\th}=\G^{\th}_{\th r}=\f{1}{r}~,~~~\G^{\th}_{\p \p}=-\sin \th \cos \th ~,\nn\\
&&\G^{\p}_{\p r}=\G^{\p}_{r \p}=\f{1}{r}~,~~~~~~~~~~~~
\G^{\p}_{\th \p}=\G^{\p}_{\p \th}=\cot \th\nn
\ea
where $\D=1-\f{2M}{r}$. For the Schwarzschild case one can put $\dot M=0$ and get the corresponding Christoffel symbols. Now, let us choose the following null tetrad  on the horizon:
\ba
\l^a=\(\pa_v\)^a+\f{\D}{2}\(\pa_r\)^a,~~n^a=-\(\pa_r\)^a,~~m^a=\f{1}{\sqrt 2r}[(\pa_{\th})^a+i\csc \th (\pa_{\p})^a],~~\bar m^a=\f{1}{\sqrt 2r}[(\pa_{\th})^a-i\csc \th (\pa_{\p})^a]\nn
\ea
such that $\l.n=-1=-m.\bar m$ is the chosen normalization and all other inner products vanish.
The nonzero components of $\cd_a\l_b$ for the Vaidya spacetime are the following :
\ba
\cd_v\l_v=-\f{\D M}{2r^2}~,~~\cd_v\l_r=\f{M}{r^2}~,~~\cd_{\th}\l_{\th}=\f{\D r}{2}~,~~\cd_{\p}\l_{\p}=\f{1}{2}\D r\sin^2\th\nn
\ea
What is interesting is that, there is no $\dot{M}$ in the expressions and one can remain assured that these above results remain unchanged for the Schwarzschild $(\dot M=0)$ case. For further clarification, if one wishes, he/she can work out the above results  beginning with constant $M$. Finally, using all the above results, it is straightforward to show that $\Th_{(\ell)}\heq 0$ irrespective of the status of $\dot M$.

For null TH, which, here, is the event horizon of Schwarzschild spacetime ( $\Phi_{00}=0$), eq.(\ref{fpb}) via eq.(\ref{fpbn}) reduces to 
\ba
\pb{R^+_{abCD}}
&\heq&-2in\pb{_{[a}\bar 
m_{b]}}o_Co_D\Psi_2+4m\pb{_{[a}\bar m_{b]}}\i_{(C}o_{D)}\Psi_2
\ea
and for non-null TH in Vaidya spacetime ($\Phi_{00}\neq 0$), eq.(\ref{fpb}) via eq.(\ref{fpbnn}) reduces to
\ba
\pb{R^+_{abCD}}
&\heq&-ih\pb{_{[a}m_{b]}}\i_C\i_D\(\Psi_2+\f{{\P}_{00}}{\mu~~}\)+4m\pb{_{[a}\bar m_{b]}}\i_{(C}o_{D)}\Psi_2+ih\pb{_{[a}\bar m_{b]}}o_Co_D
\(\f{\Psi_2}{\mu}\)
\ea
Now, we can write the equations relating the pullbacks of the bulk fields on the TH in Vaidya and Schwarzschild spacetime as follows. For the null TH in Schwarzschild spacetime the relevant equation relating the pullbacks of the bulk fields
\ba
\pb{R^+_{abCD}}
&\heq&-6in\pb{_{[a}\bar 
m_{b]}}o_Co_D\Psi_2+\Psi_2\pb{\Si_{abCD}}\label{seom}
\ea
This is precisely the equation (3.11) of \cite{ih2}  with $\Phi_{11}~\heq~0$ for Schwarzschild event horizon. Finally, the relation of the pullbacks of the relevant fields on the SSTH in Vaidya spacetime can be written as 
\ba
\pb{R^+_{abCD}}
&\heq&\Psi_2\pb{\Si_{abCD}}-ih\pb{_{[a}m_{b]}}\i_C\i_D\(3\Psi_2+\f{{\P}_{00}}{\mu~~}\)+ih\pb{_{[a}\bar m_{b]}}o_Co_D
\(\f{3\Psi_2}{\mu}\)\label{veom}
\ea
Using eq.(\ref{sigth}) in eq.(\ref{veom}) we obtain
\ba
\pb{R^+_{abCD}}
&\heq&-\(\Psi_2+\f{{\P}_{00}}{\mu~~}\)ih\pb{_{[a}m_{b]}}\i_C\i_D+\(\f{\Psi_2}{\mu}\)ih\pb{_{[a}\bar m_{b]}}o_Co_D
+4\Psi_2m\pb{_{[a}\bar m_{b]}}\i_Co_D
\label{tuitai}
\ea
Now, the scalar field $\mu$ has been  explicitly calculated in the literature\cite{booth,rezzolla} for an SSTH and it is given by
\ba
\mu=\f{T_{ab}\ell^a\ell^b}{(1/2A)-T_{ab}\ell^an^b}\label{mu}
\ea
where $T_{ab}$ is the stress-energy tensor associated with the spacetime and $A=16\pi M^2$ is the area of a 2-sphere cross-section of the SSTH.  Now, using eq.(\ref{mu}) and finding that $T_{ab}\ell^a\ell^b\heq\dot M/A, ~T_{ab}\ell^an^b\heq0$, one can calculate explicitly the value of $\mu$ on the SSTH in Vaidya spacetime  to be $2\dot M$. Then using the fact that $\Phi_{00}\heq\dot M/4M^2$ and $\Psi_2\heq-1/8M^2$ from eq.(\ref{p00}) and eq.(\ref{si2}) respectively, eq.(\ref{tuitai}) can be rewritten as
\ba
\pb{R^+_{abCD}}
&\heq&-\f{1}{16\dot MM^2}io_Co_Dh\pb{_{[a}\bar{m}_{b]}}-\f{1}{2M^2}\i_{(C}
o_{D)}m\pb{_{[a}\bar{m}_{b]}}\label{mmdot}
\ea
If the intrinsic coordinate indices of the TH be denoted by $\a,\b,\g$, 
then eq.(\ref{mmdot}) can be written as
\ba
 \e^{\a\b\g} R^+_{\a\b CD}
&\heq&\f{1}{\dot MM^2} J^{\g}_{(1) CD}+\f{1}{M^2}J^{\g}_{(2) CD}\label{csth}
\ea
where 
\ba
J^{\g}_{(1)CD}&:=&-\f{i}{16}o_Co_D\e^{\a\b\g} {h}_{[\a}{ \bar m}_{\b]},\nn\\
J^{\g}_{(2)CD}&:=&2\i_{(C}
o_{D)}\e^{\a\b\g}{m}_{[\a} {\bar m}_{\b]},\nn
\ea
$\e^{\a\b\g}$ is the volume form on the SSTH and $h,m,\bar m$ being intrinsic to the SSTH are now indexed by $\a,\b,\gamma$, etc. Eq.(\ref{csth}) is the equation of motion for a CS theory coupled to sources with dynamic couplings. While the coupling of $J_{(2)}$ is purely geometric, the coupling of $J_{(1)}$ is matter controlled as $\dot M$ is related to the energy-momentum tensor of the Vaidya spacetime. 
Thus, the nature of the SSTH is reflected by the type of external coupling to the effective CS theory on it. 


In passing, it may be noted that, if eq.(\ref{mmdot}) is pulled back to the $S^2$ i.e. a cross section of the SSTH, the equation becomes independent of the nature of the SSTH: 
\ba
\dpb{R^+_{abCD}}
&\heq&\f{1}{M^2}\dpb{\Si_{abCD}}\label{qua}
\ea
By looking at eq.(\ref{qua}) one is not able to say whether this is a field equation on a cross-section of a degenerate or non-degenerate SSTH. In fact, this equation is valid for any arbitrary 2-surface embedded in the full spacetime. Hence, quantization of eq.(\ref{qua}), as has been performed in \cite{qg1,qg2} does not uniquely specify whether it is a quantum theory of a black hole or not\footnote{This particular point has been noted earlier in \cite{viq,nor}.}. One needs to study the evolution equation of the SSTH  so as to obtain a theory of the black hole and also to probe the nature of the black hole i.e. whether it is in equilibrium or not.


\section{Conclusion}
We find that the trapping horizon in Vaidya spacetime is associated with a Chern-Simons theory coupled to two kinds of sources from the bulk geometry. While one of the coupling is purely geometric, the other one is a matter controlled coupling. This is an advancement beyond the case of equilibrium black holes whose horizon is endowed with a Chern-Simons theory coupled to bulk geometry through a constant coupling. Further, we observe that the quantization of the equation on a cross-section of the horizon is inadequate to capture the complete physics associated with the black hole horizon. A quantization of the evolution equation of the black hole horizon is at least required to make sense of the resulting theory.  

~\\
\\
{\bf Acknowledgments :} The author is indebted to Parthasarathi Majumdar for suggesting the problem and, further, for having numerous long and fruitful discussions on it. The author also wants to thank Avirup Ghosh for helping him out regarding a very crucial issue of this work. The author is indebted to an anonymous referee for giving suggestions that have drastically improved the presentation of the manuscript and also for pointing out the references\cite{booth, rezzolla}.  Parts of the work were done while the author was in Saha Institute of Nuclear Physics, Kolkata and The Institute of Mahematical Sciences, Chennai and was funded by a Doctoral fellowship and a Postdoctoral fellowship provided by the Department of Atomic Energy of the Government of India in the respective institutions. Part of the  work is funded by the DGAPA postdoctoral fellowship provided by UNAM, Mexico.

\end{document}